\documentclass[%
 aip,
 amsmath,amssymb,
 reprint,%
]{revtex4-1}

\usepackage{graphicx}
\usepackage{dcolumn}
\usepackage{bm}

\usepackage[utf8]{inputenc}
\usepackage[T1]{fontenc}
\usepackage{mathptmx}
\usepackage{xcolor}
\usepackage{tablefootnote}

\begin{document}

    \preprint{AIP/123-QED}

    \title{Strain and crystal field splitting inversion in III-Nitrides}

    \author{Fábio D. Bonani}
    \author{Anderson H. Siqueira}
    \affiliation{ 
    São Carlos Institute of Physics, University of São Paulo, 13566-590 São Carlos, São Paulo, Brazil
    }%

    \author{Horácio W. Leite Alves}
    \affiliation{%
    Departamento de Ciências Naturais, Universidade Federal de
    São João Del Rei, C.P. 110, São João Del Rei, MG 36301-160, Brazil
    }%
    
    \author{Guilherme M. Sipahi}
    \affiliation{ 
    São Carlos Institute of Physics, University of São Paulo, 13566-590 São Carlos, São Paulo, Brazil
    }%
    
    \date{\today}

    \begin{abstract}
        The wurtzite phase group III-Nitrides (AlN, GaN, InN) have attracted great interest due to their successful applications in the optoelectronics since the 90's. 
        In this paper we perform a comprehensive study of AlN, GaN and InN structural elastic and electronic properties using hybrid and conventional Density Functional Theory, presenting a comparison of the features of the three compounds. We perform a direct comparison of the features of their electronic structures, including the inversion of the top valence band associated with a negative crystal field splitting and its relation to the challenges of acceptor-doping on AlN systems.
        With the determination of elastic constants and the Young modulus we provide a simple model to connect a deformation energy associated with the parameter $u$ and the effective crystal-field splitting, showing a direct relation among internal strain and the crystal-field splitting. 
    \end{abstract}

    \maketitle


\section{\label{sec:intro}Introduction}

    The group III-Nitrides (AlN, GaN, InN) have attracted great interest due to their successful applications in the optoelectronics for the development of high luminescence efficiency, quick response time, and long lifetime blue LEDS and lasers devices, also operating in the visible and ultraviolet region \cite{nakamura2000,jain2000,Gil2013}. Their band gaps are somewhat large and direct, the main features for these type of devices, as well as for photovoltaic and thermoelectric ones \cite{lu2013} , which are still in development.

    The III-Nitrides band gap values are 0.7 eV for InN \cite{bechstedt2005,rinke2006}, 3.4 eV for GaN, and 6.2 eV for AlN\cite{nakamura2000,jain2000}. Except for InN, due to their wide band gaps and strong bond strength, these compounds can be also used for high temperature, high-frequency and highpower  transistors \cite{buniatyan2007}. However, in these devices, improvements in p-type doping are needed, due to a low hole concentration in the base at room temperature and the high access resistance \cite{kwak2002}. 

    One solution for the improvement of these electronic devices is to make uniformly doped AlGaN/GaN superlattices or interfaces. In these new complex nanostructures, there is an enhancement in hole concentration of over five orders of magnitude at reduced temperatures, as compared with to that of doped bulk samples \cite{kozodoy1999,simon2010,arakawa2016,chen2017}. However, as pointed out by Lyons and Van de Walle \cite{lyons2021}, these new techniques could work for acceptors in GaN and InN, but not for AlN, in which hole localization presents a main problem, since the self-trapped hole is a locally stable configuration in this material.

    From the theoretical side, most of the knowledge of the impurities and/or defects fingerprints on bulk materials rely on the formation energies evaluation by using supercell models together with the Density Functional Theory calculations \cite{freysoldt2014}. Within this method, transition charge levels for deep impurities were obtained and interpreted with success. E. g., due to the limitation of the computational efforts caused by their long range impurity interaction, only recently quantitative predictions of  shallow impurities features within this model were published \cite{lyons2015,lyons2021} .

    While hole traps have deep acceptor features, the hole localization is directly related to the shallow ones. Due to the effective-mass approach \cite{mireles1998} in which, together with the supercell model can give a complete understanding of this problem, the shallow levels appear as perturbed bulk valence band states. In this method, it requires a full knowledge of the wurtzite III-Nitrides band structure, specially for the valence band top states, which are the unperturbed Hamiltonian for this system. For this, Mireles and Ulloa \cite{mireles1998} have modelled the valence band states within the ${\bf k}.{\bf p}$ approach by using Rashba-Sheka-Pikus Hamiltonian.

    Due to the good results for the acceptor's energies presented by Mireles and Ulloa \cite{mireles1998}, new efforts to improve the unperturbed ${\bf k}.{\bf p}$ Hamiltonian for the wurtzite structure in order to describe, correctly, the symmetry characters, energies and wavefunctions for the top valence states of III-Nitrides as well \cite{meyer2003,rinke2008,yan2014}. 

    It is well known that the wurtzite structure (P6$_3$/mc) \cite{morkoc2000,Gil2014} is the most stable modification for the these materials. Due to the stacking features of sp$^3$ bonds in an epitaxial growth of some semiconducting materials, similar to that observed for SiC \cite{kackell1994}, the III-Nitrides can also be grown in polytypes \cite{silva2005,moriguchi2012}, but with different energetic order: the most stable structure for SiC is the 4H one instead of 2H polytype (the wurtzite structure), which is obtained for III-Nitrides. 

    Most of the polytipic structures have hexagonal and/or rhombohedric symmetries, while only one has zincblende form \cite{kackell1994}. As the wurtzite structure results from a different stacking of sp$^3$ bonds, compared to the zincblende one, we can consider that this modification is a distorted zincblende structure along its (111) direction which, in turn, reduces the symmetry of the compound from F$_{-43m}$ to P6$_3$/mc. Then, it is natural to suppose that internal strains appear along to the wurtzite {\it c} direction, even with a direct relation to the observed spontaneous polarization in these materials. 

    These internal strain effects were well exploited by Yan {\it et al.} \cite{yan2014} by improving the $6 \times 6$ ${\bf k}.{\bf p}$ Hamiltonian within the Bir and Pikus approach to model the valence band states, with good agreement with the experimental observations for InGaN alloys. However, in this new model, both spin-orbit interactions and the correct symmetry character of the highest valence band state was not taken into account, which could improve the valence band description in order to be applied in the acceptor impurities problem.

    New ${\bf k}.{\bf p}$ Hamiltonians to get the correct description of both valence and conduction bands, due to improvement of computational efforts, has been proposed recently \cite{bastos2016,bastos2018,caro2020}. Despite the fact that, the spin-orbit interactions can be considered negligible in most semiconducting materials, its inclusion helps to identify, correctly, the symmetry character of the top valence states and, thus, giving a correct description of the band states, as shown in the recent work done by Bastos and collaborators for the semiconducting zincblende structure \cite{bastos2018}. 

    Considering the improvement of ${\bf k}.{\bf p}$ modelling for wurtzite band structures, initial efforts in this direction were done by Marquardt {\em et al.} \cite{caro2020}. In their new ${\bf k}.{\bf p}$ model, more bands were evaluated from the Density Functional Theory (DFT) results (16 ones in their new model, but 8 of them were located at the conduction band) than the earlier attempts to obtain the ${\bf k}.{\bf p}$ parameters. However, the symmetry ordering for the bottom of the conduction band found in their DFT calculations is one of the possible configurations found in an earlier interpretation of resonant Raman experiments done on wurtzite GaAs \cite{kusch2012}. Moreover, there is no mention if the other configuration could be theoretically possible to obtain, or which one leads to the correct description of the GaAs ground state, as well as the importance of the internal strains effects together with the small spin-orbit perturbation can affect the conduction band states ordering. Then, the complete picture for the other semiconducting materials, which the wurtzite is an accessible form, is not fulfilled yet.

    Despite the fact that the electronic structure of III-Nitrides were extensively studied in the past, we are going to show, in this paper, from a careful analysis of DFT results of the III-Nitrides' band structures, that internal strains effects together with spin orbit effects can change the energetic order of the top valence states and, despite the fact that spin-orbit interactions are negligible in these materials, these effects combined with the internal strains in the wurtzite structure helps us to understanding the differences found for the p-type doping in AlN, when compared with GaN and InN. Together with a careful symmetry interpretation of bands, our obtained results will help on the improvements and refinements of ${\bf k}.{\bf p}$ Hamiltonians, which will be useful for the theoretical developments of future III-Nitrides based devices.

\section{\label{sec:aproach}Theoretical approach}
    
    \subsection{\label{computational}Computational details}
        The calculations involved in this work were done by using the DFT framework method within the generalized gradient approximation (GGA) proposed by Perdew–Burk–Ernzerhof\cite{perdew1996generalized} (PBE) and the hybrid functional proposed by Heyd–Scuzeria–Ernzerhof\cite{heyd2003hybrid} (HSE).
        
        To solve the Kohn–Sham equations, we have used the projected augmented wave (PAW) method\cite{blochl1994projector}, as implemented in the Vienna ab initio simulation package (VASP, version 5.4.4)\cite{kresse1993ab}, and the PAW projectors were provided by the VASP package\cite{kresse1999ultrasoft}. To describe the electronic states, the Kohn–Sham orbitals are expanded in plane waves employing a finite cutoff energy, which depends on the calculated properties. 

        We start our work with several convergence tests and we demonstrated that well converged total energies, band structures, and densities of states can be obtained by using a cutoff energy of 600 eV. Stress tensor and elastic constants calculations, however, require a cutoff energy of 800 eV, since they converge slower than other properties. 
        
        
        
        For the Brillouin zone integration, we employed a Monkhorst–Pack $\Gamma$-centered k-mesh of 11$\times$11$\times$6 to obtain the equilibrium volume, which was performed by several consecutive optimizations of the equilibrium volume to ensure that the optimized equilibrium volume is consistent with the initial set up of the basis size. On the other hand, to calculate the elastic properties, we employed a k-mesh of 22$\times$22$\times$12. 

    \subsection{\label{strain}Strain}
    
        The hexagonal wurtzite crystal structure has the symmetry defined by the space group $P6_{3}mc$  which is associated to the point group $C_{6v}$. From the symmetry group analysis, it is possible to show that it has five non-equivalent elastic constants, being they: $C_{11}$, $C_{12}$. $C_{13}$, $C_{33}$ and $C_{44}$.

        By applying the uniaxial stress–strain theory to the wurtzite crystalline phase, we have the following conditions for the deformations:
        \begin{equation}
            \sigma_{zz} = \tau; \hspace{1cm} \sigma_{xx} = \sigma_{yy} = 0  \nonumber  
        \end{equation}
        where $\tau$ is the stress. 
    
        In order to describe strain from uniaxial stress, it is necessary to determinate the Young module. This parameter is defined here as:
        \begin{equation} \label{eq:young}
            Y = \frac{\sigma_{zz}}{\epsilon_{zz}} = \frac{C_{33}(C_{11}+C_{12}) - 2C_{13}^2}{C_{11}+C_{12}}
        \end{equation}
        where $C_{11}$, $C_{12}$, $C_{13}$, $C_{33}$ are the elastic constants of the wurtzite structural phase.       
    
        Finally, considering a linear deformation, we can express the deformation energy as:
        \begin{equation}
            U = \frac{Y}{2}\epsilon^2_{zz}
            \label{eq:deformation}
        \end{equation}
    
\section{\label{sec:results}Results}

    \subsection{\label{sec:lattice} Lattice Parameters}
        The most stable crystalline phase for the majority III-V semiconductors is the zinc-blende structure. However, the III-nitrides (AlN, GaN and InN) prefer to crystallize in the hexagonal wurtzite structure, instead of the zinc-blende one\cite{Gil2013}. 
        The wurtzite structure is a hexagonal close packed lattice with four atoms per unit cell and to completely describe the crystalline structure, one needs to use three different geometrical parameters: the lattice parameters a and c, and a cell-internal structural parameter, u.

        
        \begin{table}[t]
            \caption{\label{tab:lattice} AlN, GaN and InN structural parameters, compared to experimental measures and other DFT calculations. 
            Parameters  a and c are the lattice constants (\AA) and u is the cell-internal structural parameter in the wurtzite phase.}
            \begin{ruledtabular}
                \begin{tabular}{ccccc}
                    &    & PBE & Other works & Experimental \\ \hline\hline
                    &a  & 3.113 & 3.13\footnotemark[1], 3.084\footnotemark[2] & 3.110\footnotemark[4] \\
                AlN &c & 5.013 & 5.09\footnotemark[1], 4.948\footnotemark[2] & 4.980\footnotemark[4] \\
                    &c/a & 1.610 & 1.618\footnotemark[1], 1.604\footnotemark[2] & 1.601\footnotemark[4] \\
                    &u & 0.386 & 0.382\footnotemark[1], 0.3814\footnotemark[2] & 0.382\footnotemark[4]\\
                    \hline\hline
                    &a & 3.201 & 3.24\footnotemark[1], 3.162\footnotemark[3] & 3.190\footnotemark[4] \\
                GaN &c & 5.199 & 5.24\footnotemark[1], 5.142\footnotemark[3] & 5.189\footnotemark[4] \\
                    &c/a & 1.624 & 1.614\footnotemark[1], 1.626\footnotemark[3] & 1.626\footnotemark[4] \\
                    &u & 0.379 & 0.384\footnotemark[1], 0.377\footnotemark[3] & 0.377\footnotemark[4] \\
                    \hline\hline
                    &a & 3.617 & 3.59\footnotemark[1], 3.051\footnotemark[2] & 3.5365\footnotemark[5]  \\
                InN &c & 5.825 & 5.81\footnotemark[1], 5.669\footnotemark[2] & 5.7039\footnotemark[5] \\
                    &c/a & 1.610 & 1.621\footnotemark[1], 1.619\footnotemark[2] & 1.61\footnotemark[5] \\
                    &u & 0.386 & 0.381\footnotemark[1], 0.3784\footnotemark[2] &  
                \end{tabular}
            \end{ruledtabular}
            \footnotesize{$^a$ Ref.\cite{gavrilenko2000linear}} \hspace{0.8cm} 
            \footnotesize{$^b$ Ref.\cite{wright1995consistent}} \hspace{0.8cm}
            \footnotesize{$^c$ Ref.\cite{wright1994explicit}} \hspace{0.8cm}
            \footnotesize{$^d$ Ref.\cite{schulz1977crystal}} \hspace{0.8cm}
            \footnotesize{$^e$ Ref.\cite{davydov2002absorption}} \hspace{0.8cm}
        \end{table}
        
        To determine the lattice parameters we used for the exchange potential the PBE functional and the results are summarized in the table \ref{tab:lattice}, along with other DFT calculations and experimental values.

        Starting with the AlN alloy, the calculation shows deviations from experimental values of 0.1\%, 0.66\%,  0.56\% and 1.05\% for a, c, c/a and u, respectively and are close to the parameters obtained by two other theoretical works using the PBE functional, \cite{gavrilenko2000linear}, and the LDA functional, \cite{wright1995consistent}, respectively. Our GaN results show deviations of 0.34\%, 0.19\% 0.12\% and 0.53\%, and the comparison with other theoretical results also show results close to parameters determined with PBE \cite{gavrilenko2000linear} and LDA functionals \cite{wright1994explicit}.
        Finally, our InN results show deviations of 2.28\%, 2.12\% and 0.12\%. There is no experimental measure for the u parameter in InN. Finally, the comparison with other theoretical works using PBE \cite{gavrilenko2000linear} and LDA functionals \cite{wright1995consistent}, also shows reasonable agreement.

    \subsection{\label{sec:elastic} Elastic constants}
    
        As experimental setups using III-Nitrides often present strain due to the important difference of lattice parameters, specially when InN is involved\cite{morkoc2000}
        , the determination of elastic constants is a necessary step. Five constants have to be determined for these systems: $C_{11}$ and $C_{33}$ represent the modulus for the axial compression, i.e., the stress in one direction induces a strain in the same direction; $C_{12}$ and $C_{13}$ represent the stress that induces a strain in the perpendicular directions; and, finally, $C_{44}$ represents the strain across the faces induced by the stress in a direction parallel to it.

        \begin{table}[h]
            \caption{\label{tab:constants} Elastic constants for AlN, GaN and InN calculated with PBE and HSE06 functionals, compared with other theoretical calculations and experimental values. All constants are given in GPa.}
            \begin{ruledtabular}
                \begin{tabular}{cccccc}
                &  & PBE & HSE06 & Other works\footnotemark[1] & Experimental\\ \hline\hline
                &$C_{11}$ & 376.28 & 392.39 & 410.2 & 410.5 $\pm$ 10 \footnotemark[2]\\
                &$C_{12}$ & 128.63 & 144.65 & 142.4 & 148.5 $\pm$ 10 \footnotemark[2]\\
            AlN &$C_{13}$ & 98.23 & 112.46 & 110.1 & 98.9 $\pm$ 3.5 \footnotemark[2]\\
                &$C_{33}$ & 356.59 & 366.67 & 385.0 & 388.5 $\pm$ 10 \footnotemark[2]\\
                &$C_{44}$ & 112.45 & 123.82 & 122.9 & 124.6 $\pm$ 4.5 \footnotemark[2]\\ 
                \hline \hline
                &$C_{11}$ & 329.39 & 361.117 & 368.6 & 390 \footnotemark[3] \\
                &$C_{12}$ & 127.83 & 161.509 & 131.6 & 145 \footnotemark[3] \\
            GaN &$C_{13}$ & 95.04 & 125.058 & 95.7 & 106 \footnotemark[3] \\
                &$C_{33}$ & 355.81 & 391.004 & 406.2 & 398 \footnotemark[3] \\
                &$C_{44}$ & 86.06 & 99.766 & 101.7 & 105 \footnotemark[3] \\                    
                \hline \hline
                &$C_{11}$ & 203.53 & 222.796 & 233.8 & 225 \footnotemark[4] \\
                &$C_{12}$ & 102.12 & 127.715 & 110.0 & 109 \footnotemark[4] \\
            InN &$C_{13}$ & 84.99 & 110.851 & 91.6 & 108 \footnotemark[4] \\
                &$C_{33}$ & 203.43 & 215.656 & 238.3 & 265 \footnotemark[4] \\
                &$C_{44}$ & 45.39 & 46.549 & 55.4 & 55 \footnotemark[4] \\ 
                \end{tabular}
            \end{ruledtabular}
            \footnotesize{$^a$ Ref.\cite{caro2012hybrid}} \hspace{0.8cm}
            \footnotesize{$^b$ Ref.\cite{mcneil1993vibrational}} \hspace{0.8cm}
            \footnotesize{$^c$ Ref.\cite{mcneil1993vibrational}} \hspace{0.8cm}
            \footnotesize{$^d$ Ref.\cite{serrano2011inn}} \hspace{0.8cm}
        \end{table}

    
        We determined the elastic constants using two different functionals, PBE and HSE06 and our results are summarized, along with other theoretical results and experimental values, in Table \ref{tab:constants}.
        The constants determined with the PBE functional present an average variation from the literature experimental values of 8.07\%, 13.27\% and 15.57\%, for AlN, GaN and InN, respectively. The deviation of the same constants, when calculated with the HSE06 functional, is reduced to 0.09\%, 3.04\% and 3.03\%. When compared with other HSE06  calculation~\cite{caro2012hybrid} our results show similar deviations. Analysing these deviations, one can see that, although the PBE functional may produce values closer to the experiment for $C_{13}$, HSE06 calculations clearly provide a better set of parameters.

    \subsection{\label{sec:band} Band structure}

        As pointed out in the literature, the PBE functional does not provide correctly electronic parameters of the materials, underestimating the gap energy for example. In materials with very small gap, such as the InN, there is even an inversion of bands\cite{stampfl1999density}. 
        
\begin{figure}
            \centering
            \includegraphics[scale=0.4]{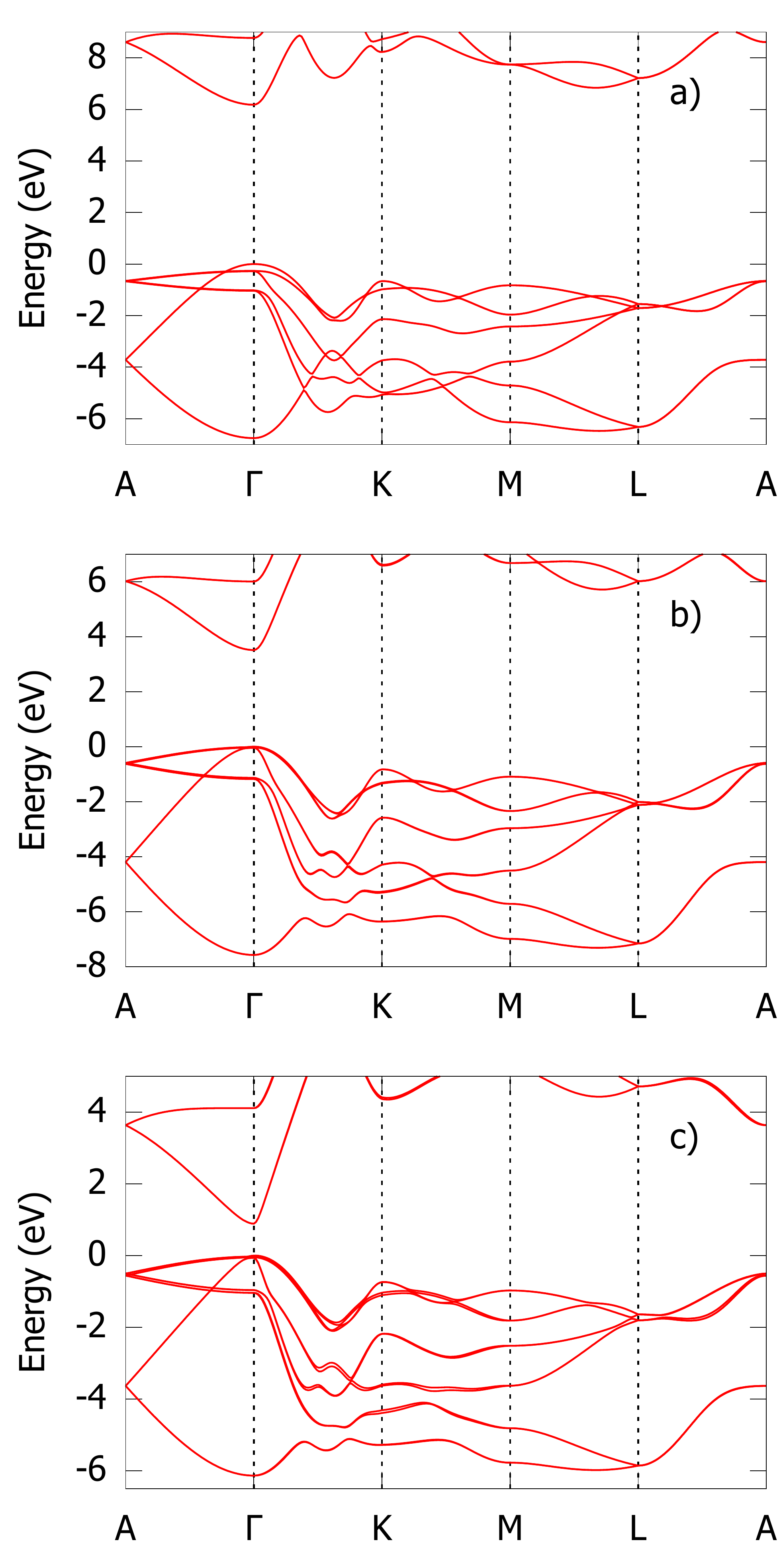}
            \caption{Band structures: a) AlN; b) GaN; c) InN.}
            \label{fig:all_complete-band}
        \end{figure}

        In order to reproduce realistic values for the band structure and the energy gap, we used the HSE$\alpha$ parametrization of the HSE functional\cite{Komsa.PhysRevB.84.075207,Colleoni.JPhysCondensMatt.28.495801}. In this approach, the mixing parameter of the PBE functional and the exact Fock exchange is defined as different from the usual 25\% value, using a linear interpolation with the PBE functional ($\alpha$=0) to approximate the experimental band gap. Even if the computational effort of this approach is even higher than the usual HSE06, since it was successfully used for zinc blend III-V systems~\cite{bastos2018}, it was our choice. We assumed values of $\alpha$: 0.355 for AlN; 0.320 for GaN and 0.305 for InN. Figure~\ref{fig:all_complete-band} presents the band structures obtained for these III-Nitrides.

\begin{figure}
            \centering
            \includegraphics[scale=0.45]{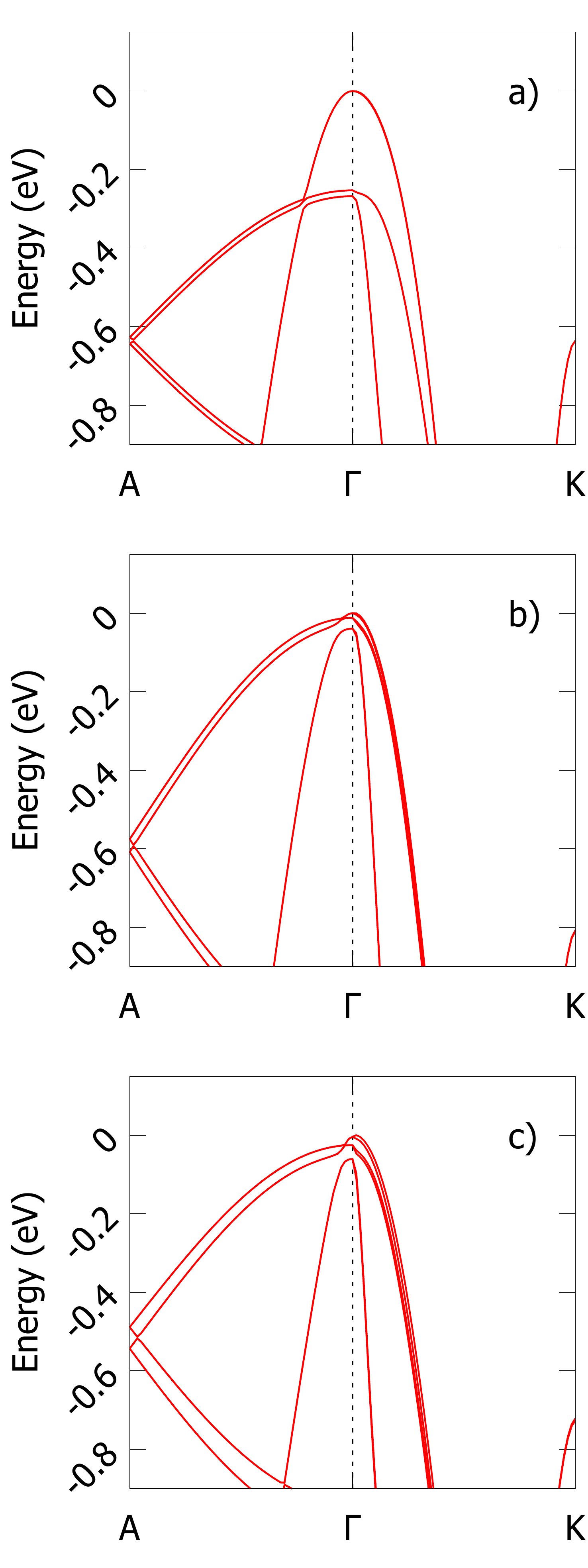}
            \caption{Band structures close to $\Gamma$ point: a) AlN; b) GaN; c) InN.}
            \label{fig:all_gamma-band}
        \end{figure}

        The overall behaviour of the different systems is very similar. The most characteristic features of these band structures is the direct band gap and the lack of band repulsion at the point A, where the bands of different symmetries cross and also the multiple crossing along the K-M direction, allowed by the symmetry analysis of the $C6_v^4$ crystal group. Apart from the overall behavior, very different features may be found when comparing the different band structures. First, there is an increase on the splitting energy among the first and second conduction bands as the atomic number of the cation increase, from AlN to InN, and second, the top of the valence band shows very different behaviors as seen in Figure~\ref{fig:all_gamma-band}.

        Looking at the valence band, its most characteristic feature is that the top of the valence band is a $\Gamma_7^\perp$ representation, followed by a $\Gamma_9$ and a $\Gamma_7^\parallel$ representations. A second very characteristic feature is the very small separation of the heavy-hole band, HH ($\Gamma_9$), band from the light hole band, LH ($\Gamma_7^ \parallel$), usually associated with the spin-orbit splitting. In wurtzite the third band is called crystal-field splitting hole band, CH ($\Gamma_7^\perp$). 
        
        The symmetry breaking fields associated with these energy splittings are the spin-orbit field, usually depicted by the $\Delta_{so}$ term, and the field  associated with the spatial break of symmetry among the planar and perpendicular directions known as the crystalline field, depicted by the term $\Delta_{cr}$. It is important to notice here that, unlike the zincblende (ZB) structures, the LH band is the spin-orbit split-off band and the CH has the combination of components usually associated with the light holes in ZB.
        
        Since the symmetry breaks of the crystalline and spin-orbit fields are  non-collinear, we need to use following expression to define the relation of the band structure splittings at $\Gamma$, $\Delta_1$ and $\Delta_2$, with the associated parameters, $\Delta_{so}$ and $\Delta_{cr}$ \cite{reynolds1996ground,de2010predicted}:
        \begin{equation}\label{eq:par}
            \Delta_{1,2} = - \frac{\Delta_{so}+\Delta_{cr}}{2} \pm \frac{\sqrt{(\Delta_{so}+\Delta_{cr})^2-u^{-1}\Delta_{so}\Delta_{cr}}}{2}
        \end{equation}
        where $\Delta_1 = \Gamma_7^\parallel - \Gamma_9$ and $\Delta_2 = \Gamma_7^\perp - \Gamma_9$. 

        Table \ref{tab:eletronic} shows the energy gaps, and parameters for the crystalline and spin-orbit field splittings for the three different alloys.
        As expected, since these calculations are done with the HSE$\alpha$ functional, the gap energies calculated for all systems have small deviations of ~1\% from the experimental data. 
        As LDA and PBE functionals are known to underestimate the gap and the HSE$\alpha$ approach adjusts it, the most effective comparison of these results with the literature, either theoretical or experimental, may be done comparing the crystal-field splitting and spin-orbit coupling parameters, $\Delta_{cr}$ and $\Delta_{so}$.
                
        \begin{table}[!htb]
            \caption{\label{tab:eletronic} Band gap ($E_g$), Spin-orbit coupling $(\Delta_{so})$ and Crystal-field-splitting $(\Delta_{cr})$ compared with measurable parameters for AlN. All energies are given in eV.}
            \begin{ruledtabular}
                \begin{tabular}{ccccc}
                    &    & HSE$\alpha$ & Other works & Experimental \\ \hline\hline
                    &$E_g$ & 6.182 &  5.8\footnotemark[1] & 6.11\footnotemark[2], 6.2\footnotemark[3], 6.28\footnotemark[4], 6.1\footnotemark[5]\\
                    AlN\ &$\Delta_{cr}$ & -0.271 &  -0.217\footnotemark[6] & -0.219\footnotemark[2] \\
                    &$\Delta_{so}$ & 0.024 & 0.019\footnotemark[6], 0.058\footnotemark[7]& 0.013\footnotemark[2] \\
                    \hline \hline
                    &$E_g$ & 3.513 & 
                    & 3.474\footnotemark[8],  3.455\footnotemark[9], 3.5031\footnotemark[11], \\
                    GaN\ &$\Delta_{cr}$ & -0.022 & 
                    0.042\footnotemark[6] & 0.022\footnotemark[8], 0.010\footnotemark[10], 0.025\footnotemark[11], \\
                    &$\Delta_{so}$ & 0.034 & 0.012\footnotemark[5], 0.013\footnotemark[6] & 0.011\footnotemark[8], 0.018\footnotemark[10], 0.017\footnotemark[11] \\
                    \hline \hline
                    &$E_g$ & 0.891 & & 0.9\footnotemark[12]
                    \\
                    InN\
                     &$\Delta_{cr}$ & -0.035 & 0.041\footnotemark[6]& \\
                    &$\Delta_{so}$ & 0.046 & 0.001\footnotemark[6] &  
                \end{tabular}
                \footnotesize{$^b$ Ref.\cite{rubio1993quasiparticle}} \hspace{0.5cm}
                \footnotesize{$^a$ Ref.\cite{li2003band}} \hspace{0.5cm}
                \footnotesize{$^c$ Ref.\cite{yim1973epitaxially}} \hspace{0.5cm}
                \footnotesize{$^d$ Ref.\cite{perry1978optical}} \hspace{0.5cm}
                \footnotesize{$^e$ Ref.\cite{vispute1995high}} \hspace{0.5cm}
                \footnotesize{$^f$ Ref.\cite{wei1996valence}} 
                \\
                \footnotesize{$^g$ Ref.\cite{suzuki1995first}} \hspace{0.5cm}
                \footnotesize{$^h$ Ref.\cite{dingle1971absorption}} \hspace{0.5cm}
                \footnotesize{$^i$ Ref.\cite{ilegems1972luminescence}} \hspace{0.5cm}
                \footnotesize{$^j$ Ref.\cite{gil1995valence}} \hspace{0.5cm}
                \footnotesize{$^k$ Ref.\cite{reynolds1996ground}} \hspace{0.5cm}
                \footnotesize{$^l$ Ref.\cite{davydov2002absorption}} 
            \end{ruledtabular}
        \end{table}

        In AlN both, the spin-orbit and the crystalline field parameter values obtained in our simulation are close to the experimental results obtained by \citeauthor{li2003band}\cite{li2003band} and close in the same range of other theoretical values with different strategies~\cite{wei1996valence,suzuki1995first}. In addition, the inversion of bands represented by the negative signs shown at the table is present in both. In GaN and InN the signal of $\Delta_{cr}$ is negative in contradiction with the theoretical work of \citeauthor{li2003band}\cite{li2003band}, although the magnitudes are of the same order. The spin orbit values are consistently bigger in our calculations than the values tabulated in their paper, 24 meV to 19 meV in AlN, 34 meV to 13 meV in GaN and 46 meV to 1 meV in InN. Experimental values for AlN and GaN  are in the same order of magnitude of the calculations.

    \subsection{\label{sec:strain} Valence band ordering and strain}
        
        The common sense, from the modeling of Wurtzite II-VI systems, puts the ordering of the valence bands from top to down being $\Gamma_9$, $\Gamma_7^\parallel$ and $\Gamma_7^\perp$.
        However, the top of valence band in AlN is known to present an inversion for a long time~\cite{li2003band}.

        \begin{figure}
            \centering
            \includegraphics[scale=0.4]{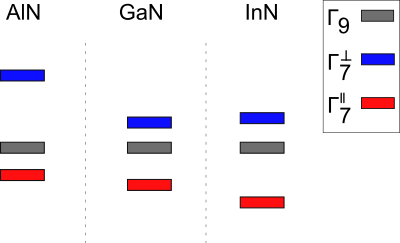}
            \caption{Band ordering of III-nitrides valence bands in the wurtzite phase.}
            \label{fig:val_ordering}
        \end{figure}

        Figure \ref{fig:val_ordering} presents a diagram of the valence band ordering of the III-nitrides whose symmetry were determined using the codes irvsp\cite{Gao2021} and IrRep, from the Bilbao Crystallographic Center\cite{Elcoro2017} with the energy differences shown in scale.
        The top valence band in all 3 cases is the usually known as CH or crystal-split hole, ($\Gamma_7^\perp$), followed by the usually known as HH or heavy hole ($\Gamma_9$) and LH or light hole ($\Gamma_7^\parallel$).
        
        It is interesting to remark, when we consider the results of the presence of the acceptors in III-Nitrides \cite{lyons2015,lyons2021}, that the CH state is the aforementioned hole trap and, its relative position to the HH and LH ones explain why is difficult to have p-type AlN related to the GaN and InN as well.
        
        To compare 
        the roles of the crystal and spin-orbit splittings we set the zero of the diagram in the $\Gamma_9$ (grey) and consequently, at $\Gamma$, $\Gamma_7^\perp$ (blue) has the value of the energy separation $\Delta_1=E_9-E_7^\perp$ (positive valued) and $\Gamma_7^\parallel$ (red) has the value of the energy separation $\Delta_1=E_9-E_7^\parallel$ (negative valued).
        This ordering leads to negative values for the crystal field parameter ($\Delta_{cr}$). Recent calculations of other III-V alloys in the wurtzite phase show $\Delta_{cr}$ with positive values \cite{de2010predicted}.

        To understand the ordering of the states it is useful to use a model based on the strain and on the differences of zincblende and wurtzite systems.
        The comparison of the WZ and ZB systems is useful here since both of them are repetitions of a layer of sp$^3$ hybridized species and the only diference between them is a rotation of 60 degrees in the subsequent layers. 
        While ZB has a repetition unit of one atom of each species, wurtzite has a unit of two atoms of each species, needing 2 different parameters to fully describe the structure  along the [0001] direction, $c$ and $u$, shown in Table ~\ref{tab:lattice}. The parameter u is related to the distance of the separation of atoms of both species in the direction of the repetition and the ideal value, defined as if the distance from one species to other were the same, is 0.375. 
        Assuming that in wurtzite, the divergence of the ideal and actual bonds is due to internal strain, one can determine an associated deformation energy associated with the crystal field splitting.


  
        \begin{table}[!htb]
            \caption{\label{tab:deformation} Calculated deformation energy U (meV), compared with eletronic parameters in eV, and Young modulus (Y) in GPa.}
            \begin{ruledtabular}
                \begin{tabular}{ccccc}
                    & AlN & GaN & InN \\ \hline
                    $\Delta_{so}$ 
                                  & 0.023 & 0.031 & 0.042  \\ 
                    $\Delta_{cr}$ 
                                  & 0.261 & 0.016  & 0.029  \\ 
                    Y
                      & 319.570 & 331.154 & 145.542 \\ 
                    U
                      & 177.333 & 13.764 & 81.061 
                \end{tabular}
            \end{ruledtabular}
        \end{table}

       Table \ref{tab:deformation} presents the deformation energy (U) and the Young modulus (Y) calculated with the HSE06 functional. We determined the Young modulus according to equation \ref{eq:young} and with the elastic constants presented at table \ref{tab:constants}.
       Our results are in good agreement with other works\cite{zagorac2019theoretical,nowak1999elastic,lu2012investigation}, presenting deviations of: 0.92\% for AlN; 1.12\% for GaN and 0.97\% for InN.

        The deformation energy was calculated using the results from the Young Modulus and assuming the elongation as the the ratio. $\epsilon_{zz}={u_{ideal}}/{u}$. The values obtained for the deformation energy are 177.3 meV, 13.7 meV and 81.1 meV, for AlN, GaN and InN, respectively. Since the spin-orbit and the strain field are not collinear, the comparison with the band structure is not straightforward. However if we use a composition of the states as a baseline, this comparison may be done. Assuming that all the states are equally important,  we set the baseline as $(E_{HH}+E_{LH}+E_{CF})/3$. Using this value, the predicted $E_{CF}$ is 250 meV, 14 meV and 78 meV, for AlN, GaN and InN, respectively.
        These values may be directly compared to $\Delta_1$ ($\Gamma_7^{\perp}-\Gamma_9)$ differences whose values, determined using HSE06, are 253 meV, 12 meV and 22 meV, for AlN, GaN and InN respectively.

        


        This simple model, based on the uniaxial strain model and applying the ratio $u_{ideal}-u/u$, relates the inversion of $\Gamma_9$ and $\Gamma_7^{\perp}$ bands, previously stated in AlN, and present also in GaN and InN alloys with the relaxation of the crystal,  expressed by the differents u. The model provides, through the sum of the deformation energy with the estimated renormalization of the baseline of the hole states, numerical values comparable to the $\Delta_1$ parameter that fit extremely well for the AlN and GaN alloys. For InN however, the comparison, although valid, gives a lesser estimation. The InN alloy, however, have a higher spin-orbit splitting that can be estimated by the separation of the $\Gamma_7^{\perp} - \Gamma_7^{\parallel}$ ($\Delta_1+\Delta_2$ in this case) and the estimation of the baseline for it is probably less accurate than in AlAs and GaAs.
    
        
    \section{\label{sec:conclusion}Conclusions}
    
In summary, we presented our theoretical results for the structural, elastic and electronic properties of wurtzite III-Nitrides in the  wurtzite crystalline structure, obtained from DFT calculations within the GGA/PBE, and the HSE approaches for the exchange-correlation functional.

Concerning the obtained structural parameters, our GGA/PBE results show good agreement with the experimental and other theoretical data. For the elastic constants results, we have observed that, by using HSE06 exchange-correlation functional there is a significant improvement of our obtained values, showing an excellent agreement with the available experimental data. 

Considering the band structures, by tuning $\alpha$ in the HSE exchange-correlation functional, better values for the band gap energies were obtained when compared to the experimental data, keeping the same energetic order of the Kohn-Sham orbitals at both the top of valence band, as well as at the bottom of the conduction band, in comparison with the HSE06 results.

Finally, we have shown, from a careful analysis of our DFT results, that internal strains effects together with spin orbit effects play a significant role in the energetic order of the top valence states. As an example, concerning to difﬁculties to obtain p-type AlN related to the GaN and InN, is directly related to the relative position of $\Gamma_7^\perp$ state (CH) compared with both the $\Gamma_9$ (HH) and $\Gamma_7^\parallel$ (LH) ones at the top of valence bands.In this case, the CH state behaves as the hole trap mentioned in previous theoretical works \cite{lyons2015,lyons2021}. We hope that our results give guidance for future experiments on this subject.

\begin{acknowledgments}
    G.M.S. acknowledges financial support from CNPq (grants No. 408916/2018-4 and 308806/2018-2) and CAPES (CsF-Grant No. 88881.068174/2014-01).
    H.W.L.A. is grateful for the financial support given by the FAPEMIG (grant No. CEX APQ 02695-14).
    A.H.S. acknowledges financial support from CAPES (grant No. 88887.371660/2019-00).
    \end{acknowledgments}
    
\nocite{*}
\bibliography{BonaniAlN}

\end{document}